\documentclass[a4paper,11pt]{article}
\usepackage{fullpage}

\usepackage{belov_paper}

\usepackage{pgfplots}

\newcommand{\smap}[1]{\stackrel{#1}{\longmapsto}}

\title{An Exponential Separation Between Quantum Query Complexity and the Polynomial Degree}
\author{Andris Ambainis\thanks{Faculty of Computing, University of Latvia, \url{ambainis@lu.lv}} 
\and
Aleksandrs Belovs\thanks{Faculty of Computing, University of Latvia, \url{aleksandrs.belovs@lu.lv}}
}
\date{}

\newcommand{\Adv}{\mathop{\mathrm{ADV}^{\pm}}}

\begin{document}

\maketitle

\begin{abstract}
While it is known that there is at most a polynomial separation between quantum query complexity and the polynomial degree for \emph{total} functions, the precise relationship between the two is not clear for \emph{partial} functions.

In this paper, we demonstrate an exponential separation between exact polynomial degree and approximate quantum query complexity for a partial Boolean function.
For an unbounded alphabet size, we have a constant versus polynomial separation.
\end{abstract}

\mycutecommand{\wdeg}{\mathop{\widetilde{\mathrm{deg}}}}

\section{Introduction}

A polynomial method is an established tool for proving lower bounds for classical~\cite{nisan:bs, nisan:pol} and quantum~\cite{beals:pol} query complexity.
In the quantum case, this method is based on an observation that a quantum query algorithm can be turned into an approximating polynomial whose degree is at most twice the query complexity of the algorithm.
Showing that a function cannot be approximated by a low-degree polynomial implies that it cannot be solved query-efficiently on a quantum computer.
This method was used early on to establish important results like the precise characterisation of quantum query complexity of total symmetric Boolean functions~\cite{beals:pol} and the optimal lower bound for the collision problem~\cite{shi:collisionLower}.
It was also used recently to prove strong lower bounds on $k$-distinctness and image size testing~\cite{bun:polynomialStrikesBack, mande:kDistinctness}.

The question of how good this lower bound technique is has gathered attention.
Nisan and Szegedy~\cite{nisan:pol} proved that $Q(f) = O\sA[\deg(f)^8]$ for total Boolean $f$, where $\deg(f)$ is the exact degree, and $Q(f)$ is the quantum query complexity.%
\footnote{
Actually, it was shown that $D(f) = O\sA[\deg (f)^8]$, where $D(f)$ is the deterministic decision tree complexity.
Here we use that $D(f)$ upper bounds approximate (and, actually, even exact) quantum query complexity.
Similar comments apply to other upper bounds below.
}
This was subsequently improved to $Q(f) = O\sA[\deg(f)^4]$ (attributed to Nisan and Smolensky in~\cite{buhrman:querySurvey}), and $Q(f) = O\sA[\deg(f)^3]$ by Midrij\= anis~\cite{midrijanis:exact}.%
\footnote{
Stated as $D(f) = O\sA[\deg(f)^4]$ and $D(f) = O\sA[\deg (f)^3]$, respectively.
}
Concerning the approximate degree $\wdeg(f)$, Beals, Buhrman, Cleve, Mosca, and de Wolf~\cite{beals:pol} showed that $Q(f) = O\sA[\wdeg(f)^6]$ for total Boolean functions.%
\footnote{
Again, stated as $D(f) = O\sA[\wdeg(f)^6]$.  It is also often cited as $D(f) = O\sA[Q(f)^6]$
}
This was improved by Aaronson, Ben-David, Kothari, Rao, and Tal~\cite{aaronson:Huang} to $Q(f) = O\sA[\wdeg(f)^4]$ for all total Boolean functions.%
\footnote{
Although the result is stated as $D(f) = O\sA[Q(f)^4]$, what is actually proven in the paper is $D(f) = O\sA[\wdeg(f)^4]$.  See also the corresponding cell in Table 1 of the paper.
}

On the other hand, Ambainis~\cite{ambainis:polVsQCC} constructed a total Boolean function with superlinear but subquadratic separation between the exact degree $\deg(f)$ and quantum query complexity $Q(f)$.
This also implies a similar separation between $\wdeg(f)$ and $Q(f)$ as $\wdeg(f)\le \deg (f)$.
Aaronson, Ben-David, and Kothari~\cite{aaronson:cheatSheets} demonstrated an almost quartic separation between $Q(f)$ and $\wdeg(f)$ as well as an almost quadratic separation between $Q(f)$ and $\deg(f)$ for a total Boolean function.
The former separation is optimal due to the aforementioned result by Aaronson \etal~\cite{aaronson:Huang}.

In order to prove a separation between quantum query complexity and the polynomial degree, one has to use a different tool than the polynomial method to prove lower bounds on quantum query complexity.
A popular alternative is the adversary method.
Indeed, Ambainis~\cite{ambainis:polVsQCC} used his, recent at the time, adversary method~\cite{ambainis:adv}.
Aaronson \etal~\cite{aaronson:cheatSheets} used their cheat sheet technique, but also relied on the lower bound for the $k$-sum problem~\cite{belovs:kSumLower}, which used the negative-weight adversary~\cite{hoyer:advNegative}, as well as other tools.

Note that all these results consider \emph{total} Boolean functions.
Up to our knowledge, the question of obtaining a separation between quantum query complexity and the polynomial degree for \emph{partial} functions has not been studied.
This is interesting, as partial functions usually allow for much larger separations.
This question was raised as an open problem in a recent survey by Aaronson~\cite{aaronson:openProblems}.

And indeed, $Q(f)$ versus $\deg (f)$ is not an exception, as we will prove the following two results in our paper (see \rf{sec:polynomials} for the precise definition of the polynomial degree):
\begin{thm}
\label{thm:main}
For every $q\ge n^{12}$, there exists a partial function $f\colon D\to\bool$ with $D\subseteq [q]^{3n}$ with the following properties:
\itemstart
\item its exact polynomial degree is at most 9;
\item its quantum query complexity is $\Omega(n^{1/3})$.
\itemend
\end{thm}

Take $q$ as the smallest power of $2$ exceeding $n^{12}$.
By replacing each variable of the function with $\log q$ Boolean variables, we get the following corollary.

\begin{cor}
\label{cor:Boolean}
There exists a family of partial Boolean functions $f\colon D\to\bool$ with $D\subseteq\cube$ satisfying the following two properties:
\itemstart
\item its exact polynomial degree is $O(\log n)$;
\item its quantum query complexity is $\widetilde\Omega(n^{1/3})$.
\itemend
\end{cor}

In \rf{sec:problem}, we formulate the problem and prove the upper bound on its polynomial degree.
In \rf{sec:lowerBound}, we prove the lower bound on its quantum query complexity.
We also use the adversary method in our proof of the lower bound.
The corresponding function is closely related to a function studied previously by Belovs and Rosmanis~\cite{belovs:tripartite}.

Finally, let us note that we have to use the negative-weight formulation of the adversary bound in our separation, and not the easier-to-apply positive-weight, which was used, for instance, by Ambainis in his aforementioned separation~\cite{ambainis:polVsQCC}.
This is because of the result by Anshu, Ben-David, and Kundu~\cite{anshu:liftingForAdversary} stating at most quadratic separation between positive-weight quantum adversary and the polynomial degree even for partial functions.

\section{Preliminaries}

For a positive integer $m$, let $[m]$ denote the set $\{1,2,...,m\}$.
Notation $\bZ_m$ denotes the additive group modulo $m$.
For a predicate $P$, we write $1_P$ to denote the indicator variable that is 1 is $P$ is true, and 0 otherwise.

For an $X\times Y$-matrix $A$, $x\in X$, and $y\in Y$, we denote by $A\elem[x,y]$ its $(x,y)$-th entry.
For $X'\subseteq X$ and $Y'\subseteq Y$, $A\elem[X', Y']$ denotes the corresponding submatrix.
Similar notation is also used for vectors.
Next, $\norm|\cdot|$ denotes the spectral norm (the largest singular value), and $\circ$ denotes the Hadamard (i.e., entry-wise) product of matrices.

We say that a linear operator $A\colon L\to K$ is an \emph{isometry from $L'$ into $K$} if its coimage is $L'\subseteq L$ and $\|Av\|=\|v\|$ for all $v\in L'$.
In other words, all the singular values of $A$ are 1, and the span of its right singular vectors is $L'$.

\subsection{Polynomials}
\label{sec:polynomials}

For a (partial) Boolean function $f\colon D \to \bool$ with $D\subseteq\cube$,  a \emph{representing polynomial} is defined as a real multivariate polynomial $P$ in variables $x_1,\dots,x_n$, treated as elements of $\bR$, such that
\itemstart
\item $P(x_1,\dots,x_n) = f(x_1,\dots,x_n)$ for every $x\in D$;
\item $0\le P(x_1,\dots,x_n) \le 1$ for all $x\in\cube$.
\itemend
The motivation behind this definition is that, as shown in~\cite{beals:pol}, every quantum query algorithm evaluating $f$ in $T$ queries exactly can be turned into a representing polynomial for $f$ of degree at most $2T$.

The notion of representing polynomial can be generalised for functions with larger input alphabets as well.
Let $f\colon D\to\bool$ with $D\subseteq [q]^{n}$ be a (partial) function with alphabet size $q$.
Then, we can define its representing polynomial (see, e.g.,~\cite{aaronson:collisionLowerOriginal})
as a polynomial in $nq$ Boolean variables $1_{x_i = a}$, where $i$ ranges over $[n]$ and $a$ over $[q]$.
Namely, it is a real multivariate polynomial $P$ satisfying the following properties:
\itemstart
\item $P(1_{x_1=1},\dots,1_{x_n=q}) = f(x_1,\dots,x_n)$ for every $x\in D$;
\item $0\le P(1_{x_1=1},\dots,1_{x_n=q}) \le 1$ for all $x\in[q]^n$.
\itemend
The motivation is similar to the Boolean case.

It may seem that the two definitions do not match for $q=2$, but in this case we have the identity $1_{x_i = 2} = 1-1_{x_i = 1}$, which allows us to remove the variables $1_{x_i = 2}$, giving essentially the same definition.

The \emph{exact polynomial degree} of a function $f$ is the minimal degree of its representing polynomial.
Similarly, one can define an approximating polynomial and the approximate degree, but we will not need these notions in the paper.

Assume for simplicity that $q=2^\ell$ is a power of two.
In this case, we can convert a function $f$ with domain in $[q]^n$ into a function $\widetilde f$ with domain in $\bool^{n\ell}$ by replacing each $a\in[q]$ by a bit-string $(a_1,\dots, a_\ell)\in \bool^\ell$ and each variable $x_i\in[q]$ with $\ell$ Boolean variables $x_{i,1},\dots,x_{i,\ell}$.
We have
\[
1_{x_i = a} = 1_{x_{i,1} = a_1}1_{x_{i,2} = a_2}\cdots 1_{x_{i,\ell} = a_\ell}.
\]
Therefore, every representing polynomial for $f$ of degree $d$ can be turned into a representing polynomial for the function $\widetilde f$ of degree $d\ell$.

\subsection{Adversary Bound}
\label{sec:adversary}
In the paper, we only use the (negative-weight) adversary bound for decision problems, which is defined as follows.

Let $f\colon D\to\bool$ with $D\subseteq[q]^n$.
An {\em adversary matrix} for $f$ is a real $f^{-1}(1)\times f^{-1}(0)$-matrix $\Gamma$.
 For any $j\in[n]$, the $f^{-1}(1)\times f^{-1}(0)$-matrix $\Delta_j$ is defined by
\begin{equation}
\label{eqn:Delta}
 \Delta_j[\![x,y]\!] = \begin{cases}0,&\text{if $x_j=y_j$;}\\1,&\text{if $x_j\neq y_j$.}\end{cases}
\end{equation}

\begin{thm}[Adversary bound~\cite{hoyer:advNegative,lee:stateConversion}]
\label{thm:adversary}
In the above notation, the quantum query complexity of the function $f$ is $\Theta\sA[\Adv(f)]$, where $\Adv(f)$ is the optimal value
of the semi-definite program
\begin{subequations}
\label{eqn:adv}
\begin{alignat}{3}
 &{\mbox{\rm maximise }} &\quad& \norm|\Gamma| \\ 
 &{\mbox{\rm subject to }} && \norm|\Delta_j\circ\Gamma|\leq 1\qquad \mbox{ for all }j\in[n] \label{eqn:advCondition}.
\end{alignat}
\end{subequations}
Here maximisation is over all adversary matrices $\Gamma$ for $f$.
\end{thm}
We can choose any adversary matrix $\Gamma$ and scale it down so that the condition
$\norm|\Delta_j\circ\Gamma|\leq 1$ holds.  Thus, we can use the condition
$\norm|\Delta_j\circ\Gamma|= \mathrm{O}(1)$ instead of $\norm|\Delta_j\circ\Gamma|\leq 1$.

Working with the matrix $\Delta_j\circ\Gamma$ might be cumbersome, so the following trick can be applied.
We write $\Gamma\smap{\Delta_j} \Gamma_j$ if $\Gamma\circ\Delta_j = \Gamma_j\circ\Delta_j$.  In other words, we modify the entries $\Gamma\elem[x,y]$ with $x_j = y_j$ to obtain $\Gamma_j$.  
As shown in~\cite{lee:stateConversion}, 
$\|\Delta_j\circ \Gamma\| \le 2\|\Gamma_j\|$, hence we can replace $\Delta_j\circ\Gamma$ with $\Gamma_j$ in~\rf{eqn:advCondition}.

\section{The Problem and the Polynomial Upper Bound}
\label{sec:problem}

The function for which we give the separation is defined in the following way.

Assume we have $3n$ input variables $x_1,x_2,\dots,x_{3n}\in [q]$.
We treat them as the elements of $\bZ_q$.
Divide the set of indices into three subsets: $A = \{1,\dots,n\}$, $B=\{n+1,\dots,2n\}$ and $C=\{2n+1,\dots,3n\}$.
Consider the following system of $n^3$ linear equations modulo $q$:
\begin{equation}
\label{eqn:system}
x_a + x_b + x_c = r_{a,b,c} ,\qquad
\text{for $a\in A$,\; $b\in B$ and $c\in C$},
\end{equation}
where $r_{a,b,c}\in\bZ_q$ are some fixed values.
We call the individual equations in~\rf{eqn:system} \emph{tripartite equations}, and the whole system of $n^3$ equations the \emph{tripartite system}.

\begin{defn}
In the \emph{threshold satisfiability problem}, given $x_1,x_2,\dots,x_{3n}$, the task is to distinguish the following two cases:
\itemstart
\item there is no equation satisfied in the tripartite system~\rf{eqn:system} (negative case); and
\item there are exactly $n$ equations satisfied in the tripartite system~\rf{eqn:system} (positive case).
\itemend
\end{defn}

Note that threshold satisfiability problem depends on parameters $r_{a,b,c}$, which are \emph{not} parts of the input, but specify a particular instance of the problem.

Although the tripartite system~\rf{eqn:system} has $n^3$ equations, the largest number of simultaneously satisfiable equations usually is much smaller.

\begin{prp}
\label{prp:numberOfEquations}
Assume $q\ge n^{12}$.
Then, there exists a choice of $r_{a,b,c}\in \bZ_q$ such that, for every input $x$, less than $4n$ of the equations in~\rf{eqn:system} are satisfied.
\end{prp}

\pfstart
This is a simple application of the probabilistic method.
In the following, all the probabilities are with respect to the uniform distribution over $r_{a,b,c}$.

Let $S$ be a subset of the equations in~\rf{eqn:system} of size $4n$.
As the number of variables is $3n$, we get that
\[
\Pr\skA[\text{All the equations in $S$ can be satisfied}] \le q^{3n-|S|} = q^{-n}.
\]
Therefore, by the union bound, the probability that it possible to satisfy at least $4n$ equations in~\rf{eqn:system} is at most 
\[
{n^3\choose 4n} q^{-n} < n^{12n} q^{-n} \le 1
\]
by our choice of $q$.
\pfend

We will call such a choice of the right-hand sides $r_{a,b,c}$ \emph{good}.

\begin{thm}
\label{thm:polynomial}
If $r_{a,b,c}$ are good, the exact polynomial degree of the threshold satisfiability function is at most $9$.
\end{thm}

\pfstart
Consider the following function
\[
K(x) = \sum_{a\in A,\; b\in B,\; c\in C} 1_{x_a + x_b + x_c = r_{a,b,c}},
\]
which counts the number of satisfied equations in~\rf{eqn:system}.
We have that
\[
1_{x_a + x_b + x_c = r} = \sum_{s,t\in \bZ_{q}} 1_{x_a = s} 1_{x_b = t} 1_{x_c = r-s-t},
\]
hence, the degree of $K$ is 3.
Take the univariate cubic polynomial
\[
T(z) = \frac14 z^3 - \frac32 z^2 + \frac94 z.
\]
The following is a plot of this polynomial.  
It has the following properties: $T(0)=0$, $T(1)=1$ and $0\le T(z)\le 1$ for all $0\le z\le 4$.
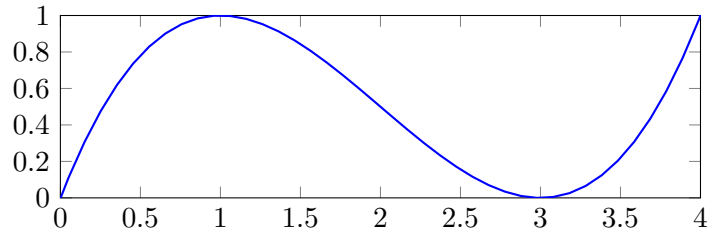
\[
\begin{tikzpicture}
\begin{axis}[xmin=0, xmax=4,ymin=0, ymax=1, width=10cm, height=4cm, samples=100]
  \addplot[blue, thick] (x, x^3/4 - 3*x^2/2 + 9*x/4);
\end{axis}
\end{tikzpicture}
\]

The polynomial $T\s[\frac{K(x)}n]$ satisfies all the requirements.
\pfend

\section{Quantum Lower Bound}
\label{sec:lowerBound}
We prove a slightly stronger result, as we prove a lower bound for an easier function.
For $s,t\in[n]$, let
\begin{equation}
\label{eqn:mu_st}
\mu_{s,t} = \sfigA{ (j,\; n+1+(j+s\bmod n),\; 2n+1+(j+t\bmod n)) \midA j\in [n] }.
\end{equation}
Each $\mu_{s,t}$ is a tripartite matching between $A=\{1,\dots,n\}$, $B=\{n+1,\dots, 2n\}$ and $C=\{2n+1,\dots,3n\}$.
We call it a shifted tripartite matching.
We use
\[
M = \{ \mu_{s,t} \mid s,t\in [n]\}.
\]
to denote the set of all shifted tripartite matchings (for a fixed value of $n$).

\begin{defn}
\label{defn:shift}
In the \emph{tripartite shift problem}, given $x_1,x_2,\dots,x_{3n}$, the task is to distinguish the following two cases:
\itemstart
\item there is no equation satisfied in the tripartite system~\rf{eqn:system} (negative case); and
\item there exists a shifted tripartite matching $\mu\in M$ such that an equation $x_a + x_b + x_c = r_{a,b,c}$ from~\rf{eqn:system} is satisfied if and only if $(a,b,c)\in \mu$ (positive case).
\itemend
\end{defn}

Since each shifted tripartite matching specifies $n$ tripartite equations, 
the tripartite shift problem is a restriction of the threshold satisfiability problem.
Therefore, any lower bound for the former is a lower bound for the latter.

A closely related problem was studied in~\cite{belovs:tripartite}.
It was like in \rf{defn:shift}, but with the following two modifications:
\itemstart
\item all $r_{a,b,c}=0$; and
\item in the positive case, it is not required that $x_a+x_b+x_c\ne r_{a,b,c}$ for $(a,b,c)\notin \mu$.
\itemend
Therefore, our result is a strengthening of~\cite{belovs:tripartite}.
We obtain a similar lower bound.

\begin{thm}
\label{thm:lower}
If $q\ge 4n^3$, the quantum query complexity of the tripartite shift problem is $\Omega(n^{1/3})$ for any choice of $r_{a,b,c}$.
\end{thm}

Essentially the same proof goes through.  
Since the differences are nonetheless substantial, we reproduce the proof in the remaining part of this section.

\paragraph{Input-Related Sets.}
We begin with defining some input-related sets.
Let 
\[
\begin{tabular}{rc@{be the set of }l}
$\tN = [q]^{3n}$ && all inputs;\\
$N$ && negative inputs; and\\
$P$ && positive inputs.
\end{tabular}
\]
For $(a,b,c)\in A\times B\times C$, let
\[
\tP^{a,b,c} = \sfigA {x\in [q]^{\{a,b,c\}} \mid x_a + x_b + x_c = r_{a,b,c}},
\]
be the solution set of the corresponding tripartite equation.
For $\mu\in M$, let
\begin{equation}
\label{eqn:Pmu}
\tP^\mu = \prod_{(a,b,c)\in \mu} \tP^{a,b,c}.
\end{equation}
In other words, $x\in[q]^{3n}$ belongs to $\tP^\mu$ if and only if all the equations of the tripartite system~\rf{eqn:system} with $(a,b,c)\in\mu$ are satisfied.
Some of the remaining equations may be satisfied as well.
Finally,
\[
\tP = \bigsqcup_{\mu\in M} \tP^\mu.
\]
We use the disjoint union here because an input $x\in [q]^{3n}$ can belong to several $\tP^\mu$ at once.
We can define $\tP$ more precisely as the set of pairs $\sfigA{ (\mu, x) \mid \mu\in M,\; x\in \tP^\mu  }$.
We consider $P$ as a subset of $\tP$, which is well-defined since $x\in P$ belongs to exactly one $\tP^\mu$.
The reason for introducing the set $\tP$ is the decomposition property~\rf{eqn:Pmu}, which $P$ lacks.

As one can guess from the notation, we use $\tN$ and $\tP$ as proxies for $N$ and $P$, respectively.
We show that their sizes do not differ too much.

\begin{clm}
\label{clm:|N|}
Under the assumption $q\ge 4n^3$, we have $|N|\ge 3|\tN|/4$ and $|P|\ge 3|\tP|/4$.
\end{clm}

\pfstart
We first prove the claim for $N$ and $\tN$.
Take $x\in \tN = [q]^{3n}$ uniformly at random.
There are $n^3$ equations in~\rf{eqn:system}.
The probability $x$ satisfies one fixed equation from this list is $1/q$.
By the union bound, the probability $x$ satisfies some equation from the list is $n^3/q \le 1/4$.
This proves $|N|\ge 3|\tN|/4$.

We can write $P = \bigsqcup_{\mu\in M} P^\mu$, where $P^\mu = P\cap \tP^\mu$ is the set of inputs satisfying precisely the equations~\rf{eqn:system} with $(a,b,c)\in\mu$.
We prove $|P^\mu|\ge 3|\tP^\mu|/4$, from which the claim follows.
To do so, we can use the same reasoning as above, because the probability a uniformly random $x\in \tP^\mu$ satisfies a fixed equation from~\rf{eqn:system} with $(a,b,c)\notin \mu$ is also $1/q$.
\pfend

\paragraph{Overview of the Proof.}
Now let us describe the general structure of the proof.
It follows the proof from~\cite{belovs:tripartite}, and is based on the ideas from~\cite{belovs:onThePower}.
The following collection $\alpha = \alpha(\mu,S)$ of real coefficients will be important:
\begin{equation}
\label{eqn:alpha}
\text{$\alpha(\mu, S)$, where $\mu\in M$ and $S\subseteq [3n]$ is such that $\absA|S\cap \{a,b,c\}|\le 1$ for all $(a,b,c)\in\mu$.}
\end{equation}
If $S$ satisfies the condition in~\rf{eqn:alpha}, we say that $S$ is \emph{good} for $\mu$.
We will implicitly assume that $\alpha(\mu, S)=0$ if $S$ is not good for $\mu$.

Let us define
\begin{equation}
\label{eqn:alphaNorm}
\norm|\alpha| = \max_{S\subseteq [3n]} \sqrt{\sum\nolimits_{\mu\in M} \alpha(\mu, S)^2 }.
\end{equation}
And, for $j\in[3n]$, we define the following operation $\partial_j$ on $\alpha$:
\[
\partial_j \alpha(\mu, S) =
\begin{cases}
\alpha(\mu, S) - \alpha(\mu, S\cup\{j\}), &\text{if $j\notin S$;}\\
0, & \text{if $j\in S$.}
\end{cases}
\]

For $\alpha$ as in~\rf{eqn:alpha}, we will define a $\tP\times \tN$ matrix $G(\alpha)$.
It satisfies the following two properties
\begin{equation}
\label{eqn:GalphaProperties}
\norm|G(\alpha)| = \norm|\alpha|
\qqand
G(\alpha) \smap{\Delta_j} G(\partial_j \alpha) \text{ for all $j\in[3n]$.}
\end{equation}
The piece of notation $\norm|\alpha|$ in~\rf{eqn:alphaNorm} was chosen precisely because of the first equation above.

We will construct an explicit $\alpha$ that satisfies the following conditions:
\begin{equation}
\label{eqn:alphaProperties}
\norm|\alpha| = n^{1/3}
\qqand
\norm|\partial_j\alpha| = O(1) \text{ for all $j\in[3n]$.}
\end{equation}

We define the adversary matrix $\Gamma$ as $G(\alpha)\elem[P, N]$.
On the one hand, $\Gamma \smap{\Delta_j} G(\partial_j \alpha)\elem[P,N]$, which has norm $O(1)$ by the above.
On the other hand, $\norm|G(\alpha)| = n^{1/3}$, and using that $P$ and $N$ are close to $\tP$ and $\tN$, respectively, we get that $\norm|\Gamma| = \Omega(n^{1/3})$.
\rf{thm:lower} follows then from \rf{thm:adversary}.

\paragraph{Fourier Basis.}
We denote $\cH = \bC^{q}$ and, for a set $T$, use notation $\cH^T = \bC^{[q]^T} = \cH^{\otimes T}$.
We often write $\cH^{a,b,c}$ instead of $\cH^{\{a,b,c\}}$ and similarly for related notions.

Let $\chi_0,\dots,\chi_{q-1}$ be the Fourier basis of $\cH$. 
Recall that it is an orthonormal basis given by $\chi_i\elem[j] = \frac1{\sqrt q}\omega_q^{ij}$, where $\omega_q = \ee^{2\pi\ii/q}$.
The most important of them is
\[
\chi_0 = \frac1{\sqrt q}\begin{pmatrix}
1\\1\\\vdots\\1
\end{pmatrix}.
\]
The Fourier basis of $\cH^T$ is given by tensor products $\chi_s=\bigotimes_{j\in T} \chi_{s_j}$, where each $s_j\in\{0,\dots,q-1\}$.
The \emph{support} of $\chi_s$ is $\{j\in T \mid s_j\ne 0\}$.

Define two orthogonal projectors in $\cH$:
\[
\Pi_0 = \chi_0\chi_0^* \qqand \Pi_1 = I - \Pi_0 = \sum_{i=1}^{q-1} \chi_i\chi_i^*.
\]
An important relation is 
\begin{equation}
\label{eqn:EDelta}
\Pi_0\smap{\Delta}\Pi_0 
\qqand
\Pi_1\smap{\Delta} - \Pi_0,
\end{equation}
where $\Delta$ is as in~\rf{eqn:Delta} and acts on the sole variable, and $\smap{\Delta}$ is consequently defined as at the end of \rf{sec:adversary}.
For $S\subseteq T$, define the projector $\Pi^T_S$ in the space $\cH^T$ by
\begin{equation}
\label{eqn:Pi^T_S}
\Pi^T_S = \bigotimes\nolimits_{j\in T} \Pi_{1_{j\in S}}.
\end{equation}
Let $\cH^T_S$ be its image.
It is equal to the span of all the Fourier basis elements of $\cH^T$ with support equal to $S$.
For a fixed $T$, the set of all $\cH^T_S$ gives an orthogonal decomposition of $\cH^T$.

We have the following properties of $\Pi^T_S$.
First, from the definition, we get the union property
\begin{equation}
\label{eqn:PiProduct}
\Pi^T_S \otimes \Pi^{T'}_{S'} = \Pi^{T\cup T'}_{S\cup S'}
\end{equation}
whenever $T$ and $T'$ are disjoint.
Next, by~\rf{eqn:EDelta}, we get the reduction property
\begin{equation}
\label{eqn:PiDelta}
\Pi^{T}_S \smap{\Delta_j} 
\begin{cases}
\Pi^T_S, & \text{if $j\notin S$;}\\
-\Pi^T_{S\setminus\{j\}}, & \text{if $j\in S$.}
\end{cases}
\end{equation}

\paragraph{The Building Blocks.}
Now let us describe the building blocks our matrices are comprised of.
Assume that $(a,b,c)\in A\times B\times C$, and $S\subset \{a,b,c\}$ is of size $|S|\le 1$.
We define
\begin{equation}
\label{eqn:Psi^abc}
\Psi^{a,b,c} _S = \sqrt{q}\, \Pi^{a,b,c}_S\elem[ \tP^{a,b,c}, {[q]^{\{a,b,c\}}} ].
\end{equation}
These are the matrices from~\rf{eqn:Pi^T_S} with $T=\{a,b,c\}$ whose rows have been restricted to the solution set $\tP^{a,b,c}$ of the corresponding tripartite equation.
The factor $\sqrt q$ is due to normalisation purposes.

\begin{clm}
\label{clm:Psi^abc}
The operator $\Psi^{a,b,c}_S$ is an isometry from $\cH^{a,b,c}_S$ into $\bC^{\tP^{a,b,c}}$.
Moreover, the operators $\Psi^{a,b,c}_\emptyset$, $\Psi^{a,b,c}_{\{a\}}$, $\Psi^{a,b,c}_{\{b\}}$, and $\Psi^{a,b,c}_{\{c\}}$ have pairwise orthogonal ranges.
\end{clm}

\pfstart
From the definition, it is clear that the coimage of $\Psi^{a,b,c}_S$ is contained in the coimage of $\Pi^{a,b,c}_S$, which is $\cH^{a,b,c}_S$.
The operators 
$\Psi^{a,b,c}_\emptyset$, $\Psi^{a,b,c}_{\{a\}}$, $\Psi^{a,b,c}_{\{b\}}$, and $\Psi^{a,b,c}_{\{c\}}$
map the corresponding Fourier basis elements
\[
\chi_0\otimes \chi_0\otimes \chi_0, \quad
\chi_{s_a}\otimes \chi_0\otimes \chi_0, \quad
\chi_0\otimes \chi_{s_b}\otimes \chi_0, \quad
\chi_0\otimes \chi_0\otimes \chi_{s_c},
\]
where $s_a, s_b, s_c$ are non-zero, into the vectors
\begin{equation}
\label{eqn:basisImages}
\begin{aligned}
\sqrt q (\chi_0\otimes \chi_0\otimes \chi_0)\elem[\tP^{a,b,c}], \quad
&&\sqrt q (\chi_{s_a}\otimes \chi_0\otimes \chi_0)\elem[\tP^{a,b,c}],\\
\sqrt q (\chi_0\otimes \chi_{s_b}\otimes \chi_0)\elem[\tP^{a,b,c}], \quad
&&\sqrt q (\chi_0\otimes \chi_0\otimes \chi_{s_c})\elem[\tP^{a,b,c}],
\end{aligned}
\end{equation}
respectively.
It remains to prove that all these vectors together form an orthonormal system in  $\bC^{\tP^{a,b,c}}$.

We can identify $x\in \tP^{a,b,c}$ with $x\in [q]^{\{a,b\}}$ as the third element $x_c$ is uniquely determined by $x_c = r_{a,b,c} - x_a - x_b$.
Therefore, we may treat the vectors from~\rf{eqn:basisImages} as belonging to $\cH^{a,b}$.
Under this assumption, the first three vectors in~\rf{eqn:basisImages} become
\begin{equation}
\label{eqn:basisImagesA}
\chi_0\otimes \chi_0,\qquad
\chi_{s_a}\otimes \chi_0,\qqand
\chi_0\otimes \chi_{s_b}.
\end{equation}
(Here we used the $\sqrt{q}$ prefactor to compensate for one missing $\chi_0$.)
Considering the last vector, its entry corresponding to $x\in \tP^{a,b,c}$ is
\[
\tfrac1q \omega_q ^{s_cx_c} = 
\tfrac1q \omega_q ^{s_c(r_{a,b,c} - x_a - x_b)} = 
\tfrac{\omega_q^{s_cr_{a,b,c}}}q \omega_q ^{-s_cx_a - s_cx_b}.
\]
Hence, the last vector of~\rf{eqn:basisImages} becomes
\begin{equation}
\label{eqn:basisImagesB}
\omega_q^{s_cr_{a,b,c}}\; \chi_{-s_c}\otimes \chi_{-s_c}.
\end{equation}
Clearly, the vectors in~\rf{eqn:basisImagesA} and~\rf{eqn:basisImagesB} form an orthonormal system.
\pfend

Now, let $\mu\in M$, and $S\subseteq [3n]$ be good for $\mu$, i.e., $|S\cap\{a,b,c\}|\le 1$ for every $(a,b,c)\in \mu$.
We define the operator
\begin{equation}
\label{eqn:Psi^mu_S}
\Psi^\mu_S 
= q^{n/2} \Pi^{[3n]}_S \elem[\tP^\mu, \tN]
= \bigotimes_{(a,b,c)\in \mu} \Psi^{a,b,c}_{S\cap\{a,b,c\}},
\end{equation}
where the equality follows from the union property~\rf{eqn:PiProduct} and the definition~\rf{eqn:Pmu} of $\tP^\mu$.

\begin{clm}[Orthogonal Isometry Claim]
\label{clm:Psi^mu}
The operator $\Psi^\mu_S$ is an isometry from $\cH^{[3n]}_S$ into $\bC^{\tP^\mu}$.
Moreover, for a fixed $\mu$, the ranges of $\Psi^\mu_S$ are pairwise orthogonal.
\end{clm}

\pfstart
This follows from~\rf{clm:Psi^abc} and the second definition of $\Psi^\mu_S$ in~\rf{eqn:Psi^mu_S}.
\pfend

Also, from~\rf{eqn:PiDelta} and the first definition of $\Psi^\mu_S$ in~\rf{eqn:Psi^mu_S}, we get the reduction property
\begin{equation}
\label{eqn:PsiDelta}
\Psi^\mu_S \smap{\Delta_j} 
\begin{cases}
\Psi^\mu_S, & \text{if $j\notin S$;}\\
-\Psi^\mu_{S\setminus\{j\}}, & \text{if $j\in S$.}
\end{cases}
\end{equation}

\paragraph{The Matrix $G(\alpha)$.}
Now we are able to define the matrix $G(\alpha)$ for $\alpha$ from~\rf{eqn:alpha}.
It is a $\tP\times \tN$-matrix defined as the vertical stack of matrices
\begin{equation}
\label{eqn:Galpha}
G(\alpha) = 
\begin{pmatrix}
G^{\mu_{1,1}}(\alpha)\\\vdots\\ G^{\mu_{n,n}}(\alpha)
\end{pmatrix},
\end{equation}
where we have one block
\[
G^\mu(\alpha) = \sum_{S\subseteq[3n]} \alpha(\mu, S) \Psi^\mu_S
\]
for each shifted tripartite matching $\mu\in M$.
Recall that we implicitly assume that $\alpha(\mu, S)=0$ if $S$ is not good for $\mu$, therefore, the $\Psi^\mu_S$ that appear in the latter sum are well-defined (satisfy the conditions above~\rf{eqn:Psi^mu_S}).

For two $\alpha$ and $\alpha'$ as in~\rf{eqn:alpha}, we can define $\alpha+\alpha'$ element-wise, and $G(\alpha)$ is linear in $\alpha$: $G(\alpha+\alpha') = G(\alpha) + G(\alpha')$.

\begin{clm}
\label{clm:GalphaProperties}
Thus defined matrix $G(\alpha)$ satisfies the claims in~\rf{eqn:GalphaProperties}: $\norm|G(\alpha)| = \norm|\alpha|$
and 
$G(\alpha) \smap{\Delta_j} G(\partial_j \alpha)$ for all $j\in[3n]$.
\end{clm}

\pfstart
Let us start with the first claim.
We can write $G(\alpha) = \sum_S G(\alpha|_S)$, where 
\begin{equation}
\label{eqn:alpha|S}
\alpha|_S(\mu, S') = 
\begin{cases}
\alpha(\mu, S'),& \text{if $S'=S$}\\
0, &\text{otherwise}.
\end{cases}
\end{equation}
We have that $G^\mu(\alpha|_S) = \alpha(\mu, S) \Psi^\mu_S$.
Since by the Orthogonal Isometry \rf{clm:Psi^mu}, each $\Psi^\mu_S$ is an isometry from $\cH^{[3n]}_S$, we get
\[
\normA| {G(\alpha|_S)}| = \sqrt{\sum\nolimits_{\mu\in M} \alpha(\mu, S)^2 }.
\]
By the same \rf{clm:Psi^mu}, the ranges and coimages of all $G(\alpha|_S)$ are pairwise orthogonal.
Hence, $\norm|G(\alpha)| = \max_S \normA| {G(\alpha|_S)}| = \norm|\alpha|$ as defined in~\rf{eqn:alphaNorm}.

By the reduction property~\rf{eqn:PsiDelta}, the second claim holds for each block of $G(\alpha)$, that is:
\[
G^\mu(\alpha) \smap{\Delta_j} G^\mu(\partial_j \alpha).
\]
Therefore, it holds for the whole matrix $G(\alpha)$.
\pfend

\paragraph{Construction of $\alpha$.}
We define
\begin{equation}
\label{eqn:alphacMS}
\alpha(\mu, S) = \frac1n \max\sfig{n^{1/3} - |S|, 0}
\end{equation}
if $S$ is good for $\mu$.  Otherwise, we assume $\alpha(\mu,S)=0$.

\begin{clm}
\label{clm:alphaProperties}
The $\alpha$ as defined in~\rf{eqn:alphacMS} satisfies the conditions in~\rf{eqn:alphaProperties}: 
$\norm|\alpha| = n^{1/3}$
and
$\norm|\partial_j\alpha| = O(1)$ for all $j\in[3n]$.
\end{clm}

\pfstart
The value $\|\alpha\|=n^{1/3}$ is attained at $S=\emptyset$.

Now let us prove the second property.
It suffices to check that, for all $S$ and $j\in [3n]\setminus S$,
\begin{equation}
\label{eqn:alphaOne}
\sum\nolimits_{\mu\in M} \sA[\alpha(\mu, S) - \alpha(\mu, S\cup\{j\}) ]^2  = O(1).
\end{equation}
Fix $S$ and $j$.  If $|S|\ge n^{1/3}$, then~\rf{eqn:alphaOne} is zero, so we may assume $|S|\le n^{1/3}$.
There are $n^2$ choices of $\mu\in M$.  They fall into three categories:
\begin{itemize}
\item 
$S$ is not good for $\mu$.  Then, $S\cup\{j\}$ is also not good for $\mu$ and
\[
\alpha(\mu, S) - \alpha(\mu, S\cup\{j\}) = 0.
\]

\item 
$S\cup \{j\}$ is good for $\mu$.  Then, $S$ is also good for $\mu$, and
\[
\absA|\alpha(\mu, S) - \alpha(\mu, S\cup\{j\})| \le \frac 1n.
\]
(The $\le$ case may hold for $|S| = \floor[n^{1/3}]$. Otherwise, we have equality.)

\item
$S$ is good for $\mu$, but $S\cup\{j\}$ is not good for $\mu$.
In this case, we have an upper bound of
\[
\absA|\alpha(\mu, S) - \alpha(\mu, S\cup\{j\})|  = 
\absA|\alpha(\mu, S)|  \le
n^{-2/3}.
\]

\end{itemize}

Let us estimate for how many $\mu$ the third option above holds.
This happens only if there is $(a,b,c)\in \mu$ such that $j\in \{a,b,c\}$ and $\absA|S\cap \{a,b,c\}| = 1$.
Assume for concreteness $j = a\in A = \{1,\dots,n\}$, the other two cases being similar.
Let $\mu = \mu_{s,t}$ as defined in~\rf{eqn:mu_st}.
We get that the third option holds only if
\[
\text{either\quad $n+1+(j+s\bmod n)$\quad or\quad $2n+1+(j+t\bmod n)$\quad belongs to $S$.}
\]
There are at most $|S|n \le n^{4/3}$ choices of $s$ and $t$ that satisfy the above condition.
Therefore,
\[
\sum_{\mu\in M} \sA[\alpha(\mu, S) - \alpha(\mu, S\cup\{j\})]^2
\le n^2\cdot \frac1{n^2} + n^{4/3}\cdot n^{-4/3} = O(1). \qedhere
\]
\pfend

\paragraph{Finishing the Proof}
As mentioned previously, we define the adversary matrix as
\[
\Gamma = G(\alpha) \elem[P,N]
\]
with the choice of $\alpha$ from~\rf{eqn:alphacMS}.
Then, $\Gamma\smap{\Delta_j} G(\partial_j \alpha)\elem[P,N]$ and for the latter matrix by~\rf{eqn:GalphaProperties} and~\rf{eqn:alphaProperties}, we have
\[
\normA| G(\partial_j \alpha)\elem[P,N] | \le
\normA| G(\partial_j \alpha) | =
\norm| \partial_j \alpha | =
O(1).
\]

It remains to lower bound $\|\Gamma\|$.
We can write $\alpha = \alpha' + \alpha''$ where $\alpha' = \alpha|_\emptyset $ as in~\rf{eqn:alpha|S} and $\alpha'' = \alpha - \alpha'$.
Let $u_P$, $u_{\tP}$, $u_N$, and $u_{\tN}$ denote the uniform unit vectors in $\bC^P$, $\bC^{\tP}$, $\bC^N$ and $\bC^{\tN}$, respectively.
That is, $u_P\elem[x] = 1/\sqrt{|P|}$ for all $x\in P$, and similarly for other vectors.
We have
\begin{equation}
\label{eqn:Estimate1}
\norm|\Gamma| \ge 
u_P^* \Gamma u_N =
u_P^* \sA[{G(\alpha') \elem[P,N]}] u_N + 
u_P^* \sA[{G(\alpha'') \elem[P,N]}] u_N.
\end{equation}
We bound both terms separately.

By construction, $G(\alpha')$ is an $\tP\times \tN$ matrix which is a vertical stack of matrices~\rf{eqn:Galpha}, where each block is $n^{-2/3} \Psi^\mu_\emptyset$.
Each $\Psi^\mu_\emptyset$ has all its entries equal (to $q^{-5n/2}$).
Thus, 
\[
u_{\tP}^* G(\alpha') u_{\tN} = \|G(\alpha')\| = \|\alpha'\| = n^{1/3}.
\]
Using~\rf{clm:|N|}, we get
\begin{equation}
\label{eqn:Estimate2}
u_P^* \sA[{G(\alpha') \elem[P,N]}] u_N
 = \sqrt{\frac{|P|\cdot|N|}{|\tP|\cdot |\tN|}}\; u_{\tP}^* G(\alpha') u_{\tN}
\ge \frac{3n^{1/3}}4.
\end{equation}

Now consider the second term.
First, we have
\[
\norm| G(\alpha'') | = \norm |\alpha''| < n^{1/3}.
\]
Next, by \rf{clm:Psi^mu}, the vector $u_{\tP}$ is orthogonal to the range of $G(\alpha'')$.
Therefore,
\[
u_{\tP}^* G(\alpha'')\elem[\tP, N] = 0.
\]
\mycutecommand{\wu}{\widetilde u}
We will use the vector $\wu =  \sqrt{|\tP|/|P|} u_{\tP}$.
It has the property $\wu\elem[P] = u_P$.
Also, by \rf{clm:|N|}, $\normA|\wu\elem[\tP\setminus P]| \le 1/\sqrt3$.
Thus,
\begin{equation}
\label{eqn:Estimate3}
\begin{aligned}
u_P^* \sA[{G(\alpha'') \elem[P,N]}] u_N
&= \wu^* \sA[{G(\alpha'') \elem[\tP,N]}] u_N - \wu\elem[\tP\setminus P]^*  \sA[{G(\alpha'') \elem[\tP\setminus P,N]}] u_N \\
&\ge - \normA |\wu\elem[\tP\setminus P]| \cdot \normA| G(\alpha'') \elem[\tP\setminus P,N] |
\ge - n^{1/3}/\sqrt3.
\end{aligned}
\end{equation}

Combining Eq.~\rf{eqn:Estimate1}, \rf{eqn:Estimate2} and~\rf{eqn:Estimate3}, we get
\[
\|\Gamma\| \ge \sB[\frac 34 - \frac 1{\sqrt{3}}] n^{1/3} = \Omega(n^{1/3}).
\]

\section{Discussion}
The choice of the problem~\rf{eqn:system} has been chiefly motivated by the availability of a relatively simple lower bound in~\cite{belovs:tripartite}.
In principle, it is possible to analyse other problems.
For instance, consider a problem of the form
\begin{equation}
\label{eqn:simpler}
x_a + x_b = r_{a,b} ,\qquad\text{for $a\ne b$ in $[2n]$,}
\end{equation}
and take $M$ as consisting of all perfect matchings in $[2n]$.

If $+$ is the bit-wise xor and all $r_{a,b}$ are zeroes, we get the collision problem.
The lower bound on its quantum query complexity is $\Omega(n^{1/3})$ and it was obtained by Aaronson and Shi~\cite{shi:collisionLower} using the polynomial method.
We see no reason to expect the non-homogeneous case (with $r_{a,b}$ non-zero) to be any simpler, but the argument of \rf{sec:problem} shows that it completely breaks down Aaronson's and Shi's lower bound.
For the adversary method, which we used in \rf{sec:lowerBound}, there was essentially no difference between the homogeneous case of~\cite{belovs:tripartite} and the non-homogeneous case of the current paper.
The right-hand sides $r_{a,b,c}$ of~\rf{eqn:system} manifested themselves as the phases in~\rf{eqn:basisImagesB}, which are irrelevant to the proof.
It would be interesting to establish lower bounds for the problem in~\rf{eqn:simpler}.  This would improve the constants in our lower bounds.

Concerning \rf{cor:Boolean}, it is not clear whether $O(\log n)$ can be improved to something better, thus resulting in a superexponential separation.

\subsection*{Acknowledgements}
We thank Scott Aaronson for writing the open problem survey \cite{aaronson:openProblems}
which attracted our attention to this problem.
We also thank the anonymous reviewers at the CCC conference for their numerous valuable suggestions on the presentation of the paper.

This work has been supported by the ERDF project number 1.1.1.5/18/A/020 ``Quantum algorithms: from complexity theory to experiment.''

\bibliographystyle{habbrvM}
{
\small
\bibliography{belov}
}

\end{document}